# PHENOMENOLOGICAL MODEL FOR GROWTH OF VOLUMES DIGITAL DATA

Andrey V. Makarenko, Ph. D.

R & D Group "Constructive Cybernetics", Russia

e-mail: avm@rdcn.ru

**Abstract.** Currently, experts from IT industry are closely monitoring the soaring total volume of digital data. Moreover the problem is not purely technical, it directly affects human civilization as a whole. The growth rate of the all increasing and is already very large. Began is actively appear apocalyptic scenarios of development IT technology, and humanity as a whole. In this paper we propose a constructive alternative to these emotional ideas. Invited to consider the digital industry as a complete system that is developing in close connection with human civilization. Moreover, system self-organizing and essentially nonlinear in its behavior. To study this system is applied system-cybernetic approach. The mathematical model is developed, shows that in the future rate of production of digital data is stabilize at 13.2 ZB per year.

**Keywords:** simulation, growth of volumes, digital data, forecast, singularity, systems approach.

## Introduction

Impetuous growth of production volumes of digital data – is one of the major problems of the present information technology. It a creates the several of difficulties associated with the storage, transmission and processing of data [1]. Furthermore, huge volumes of data the hard to manage. Consequently, the cost of IT-systems increases, a their effectiveness reduced.

The main factor contributing to the growth of production volumes of digital data - it a total penetration of digital technologies in the life of every person on Earth [2]. Have appeared even such terms as: "Digital Universe", "Digital Life", "Digital Shadow", "Digital Footprint" [3]. Increasing volumes of data that are directly related to the life of a particular person, creates three major problems:

- protection data from unauthorized access;
- maintaining the confidentiality of private life;
- information overload of sensory and nervous systems of human.

Thus, the problem of growth of production volumes of digital data, from the a purely technical, is transformed into social problem. Such positioning makes the task of estimating and forecasting the growth of volumes of digital data even more urgent.

In the description and study of this process is usually applied a detailed analytical approach [2]. In this case, digital information is represented as the sum of the individual components belonging to different technologies and areas of human life. Each piece of data and the group is regarded as a subsystem, which follows its own path of development. For the prediction of the future, these partial components continues beyond the present. Is natural, they follow the statistical laws. It should be noted that the statistical information (see eg review in [2]) this key to understanding the laws are suitable for quantitative description of the dynamics of growth of production volumes of digital data. All this makes it possible predict the total amount of digital data, who is creates a humanity in the future. This problem is posed and solved in the present paper. In the solution this problem involved methods of systems analysis and synergetics. They are specifically created is for a phenomenological description and study of complex systems.

## 2. Source data and staging the problem

Time Now experts of the IT industry closely following the rapid growth in aggregate output of volumes digital data. The growth rate of all the increases and already very large (see Table 1). It began be perceived as an information explosion that is able to shake humanity.

**Table 1**
Total amount of digital data generated for per year (*)

| Year | 2006 | 2007 | 2008 | 2009 | 2010 | 2011 | 2012 |
|---|---|---|---|---|---|---|---|
| Volumes Data, EB | 185.62 | 319.71 | 486.52 | 762.89 | 1143.23 | 1699.48 | 2502 |

\* – data in Table 1 are obtained by combining the information given in [1-3].

Naive extrapolation into the future, lead to a disappointing forecasts and even the apocalyptic scenarios of development of IT technologies (theory of technological and information of the singularity [4, 5]), see Figure 1.

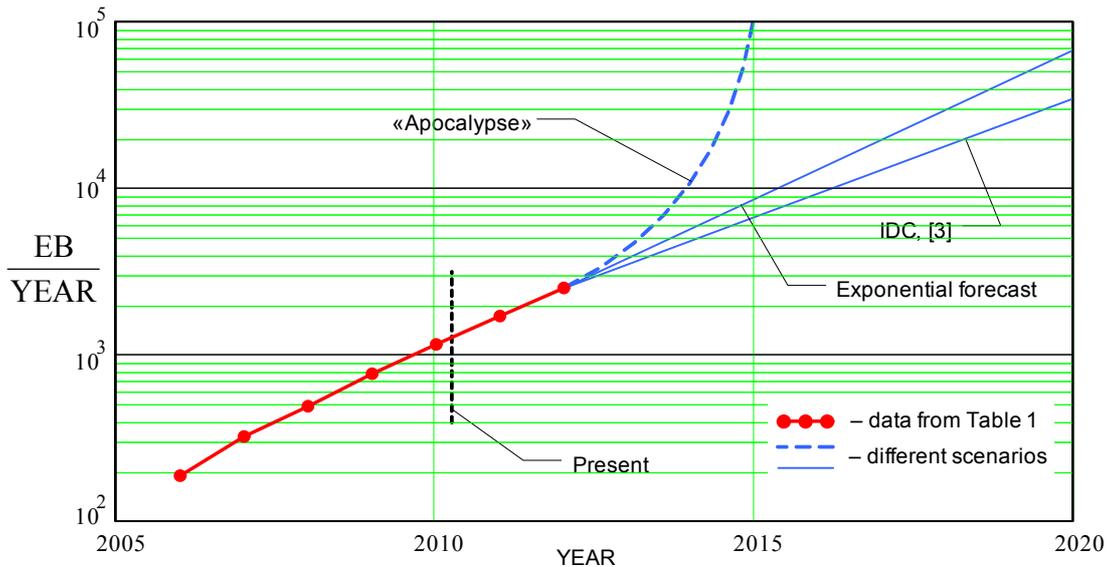

**Figure 1.** Various scenarios growth of production volumes a digital information

In these circumstances, is necessary to seek alternative approaches to the study of the dynamics of growth of volumes of information content. This will facilitate a more detailed understanding of patterns of evolution of the digital industry. Constructive alternative to the apocalyptic scenario [4, 5], is this a system-cybernetic approach to the digital industry. As an integral system that develops in close connection with human civilization. Moreover, self-organizing system and essentially nonlinear in its behavior. This concept is the basis of a mathematical model, which is discussed below. This model, to describe the dynamics of growth of volumes digital data, uses these stand.

### 3. A mathematical model is production volumes of digital data

Total production of digital data for the year at the moment time $T$ we shall characterized by the value $V(T)$. Where the time $T$ is expressed in years, $T \geq 0$, while the volume $V$ – in exabytes (EB, $\times 10^{18}$). The physical essence of $V$ defines the two principal requirements:
$$V(T) \geq 0 \ \forall T, \ \lim_{T \to 0} V(T) = 0. \qquad (1)$$
The first requirement reflects a non-negativity of volumes data. Second – the lack of digital technology at the dawn of the appearance of humanity. To construct the model there are 7 equidistant sampling of time series (Table 1). Initial data denote $\hat{V}$.

Judging from the nature of the data a simple phenomenological model of the process can be an exponential function:
$$V(T) = a e^{b(T-T_0)} + c, \qquad (2)$$
Here and below $T_0$ – constant characterizing the turning point a situation (volume of digital data has exceeded volume the analog data) [2], $T_0 = 2002$ год. At the same time based on (1), in (2) should be the value $c = 0$. According to the Table 1, by LSM defines the following values of the coefficients (2): $a = 44.417592$, $b = 0.403777$. Correlation Pearson coefficient $\text{Corr}(V(T), \hat{V}) = 0.9996932$. Applying to the model (2) test on the prediction of the "backward". We calculate the moment of time:

$$T_b = \left(T : V(T) = \frac{1}{8 \cdot 10^{18}}\right). \tag{3}$$

Value $T_b$ – this is the moment of time when the production of digital data reached a rate of 1 bit per year. Model (2) gives an estimate $T_b \approx 1884.81$. Compare this number with the historical data that describe the emergence of the digital age[1] [6-10]:

- 1832 – electromagnetic telegraph, Baron von Schilling-Canstadt;
- 1848 – upgraded version of Morse code, Samuel Morse, Alfred Vail, Friedrich Clemens Gerke;
- 1850 – direct-printing telegraphy, Boris Semyonovich von Jacobi;
- 1869 – universal ticker machine, Thomas Edison;
- 1870 – Baudot code, Jean-Maurice-Émile Baudot;
- 1920 – a global teleprinter network "Telex network";
- 1932 – international standard for telegraphic code ITA2 (CCITT-2);
- 1943 – computer "COLOSSUS", Alan Turing; software-driven computer "MARK-1", Howard Aiken, IBM.

Naive averaging given dates gives the result – 1883 year. Empirical estimation is contained in [5] – 1900 year. As a first approximation in test (3) is model (2) gives the weighted average absolute error $\Delta_{Tb} \approx 0.905$ years. The relative error (specific to forecast period), equal $\Delta^*_{Tb} \approx 0.747$ %.

Thus, the model (2) shows a high correlation coefficient with the real data $\hat{V}$. And the passes an effective test on the prediction of the "backward". Negative property of the model (2) – is an unlimited increase in the $V(T)$. If we assume that model (2) is true, then in 2342, the amount of information will be created (in bits), which will exceed number of elementary particles in the universe (upper estimate $\sim 10^{80}$ [11]). This is the strongest argument against the supporters of unrestricted exponential growth [5]. In reality, this certainly could not be. Information is required material carrier. Consequently, the model (2) does not adequately reflect the future the process generation of digital content. Is possible to correct the situation by introducing an additional requirement:

$$\lim_{T \to +\infty} V(T) << \infty. \tag{4}$$

Constraints (1) and (4) require the asymptotic conduct of the function $V(T)$. It condition meets the logistic curve:

$$V(T) = \frac{a}{1 + b e^{-c(T - T_0)}}. \tag{5}$$

According to the Table 1, by LSM defines the following values of the coefficients (5): $a = 1.320155 \cdot 10^4$, $b = 390.843634$, $c = 0.451372$. Correlation Pearson coefficient $\text{Corr}(V(T), \hat{V}) = 0.9999361$. This value is greater than at model (2). Applying to the model (5) test on the prediction of the "backward" (3). It gives an estimate $T_b \approx 1897.78$. The errors have the values: $\Delta_{Tb} \approx 7.39$ years and $\Delta^*_{Tb} \approx 6.829$ %. The magnitude errors more than the model (2), but they are quite acceptable. Note that the initial data (see Table 1) themselves contain errors and inaccuracies.

---

[1] We consider is exactly of electrical carrier of information. Other carriers of digital information has already been used previously [6-10]. In 1642 Blaise Pascal presented the "Pascaline" – the first really well-known mechanical digital computing device. In 1673 the mathematician Gottfried Wilhelm Leibniz created a binary mechanical calculator. Punch cards were first used in the Jacquard loom in 1808 to control patterns on tissues.

Thus, a model (5) limits the value of $V(T)$:
$$V_\infty = \lim_{T \to +\infty} V(T) \approx 13.2 \text{ ZB}.$$

## 4. Results of simulation

Behavior value of $V(T)$, calculated using the model (5), is shown on Figure 2.

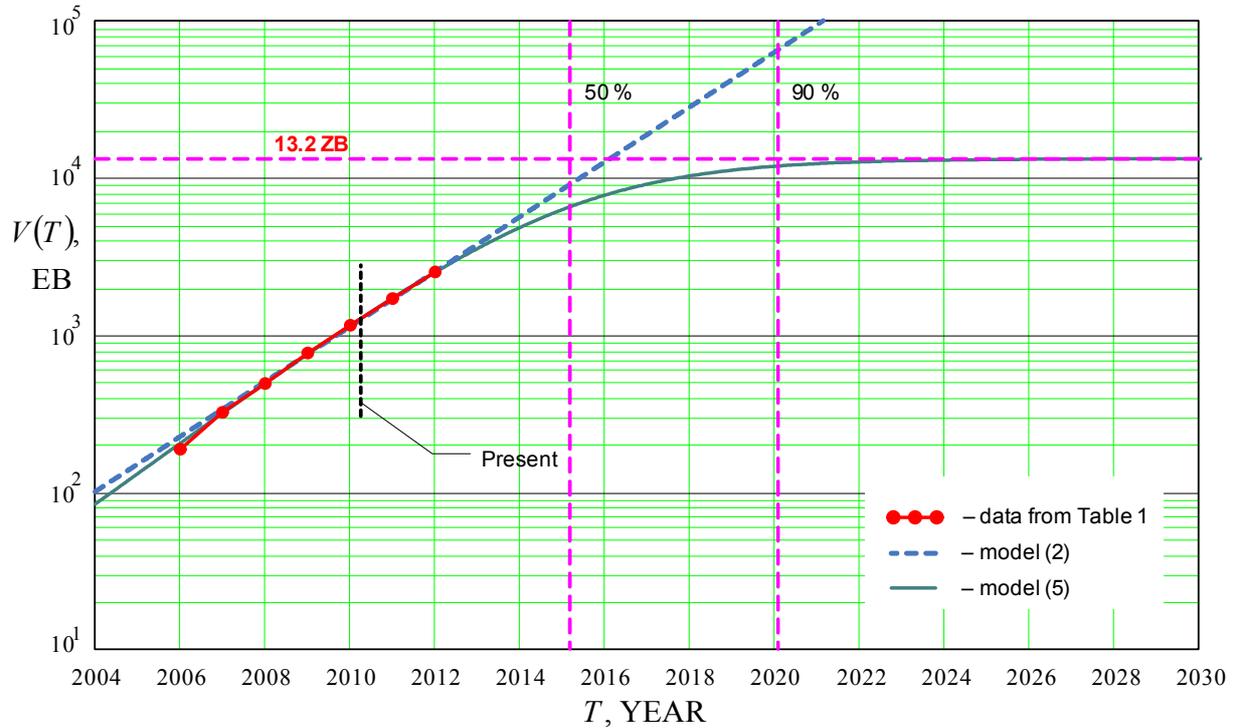

**Figure 2.** Forecast of production growth a volumes digital data, model (5)

The main a consequence of the model (5): in the future production volume of digital data is stabilize at 13.2 ZB per year. Level of 50 %, from the limiting value $V_\infty$, humanity will reach by mid 2015. Level 90 % – by 2020. By mid 2011, humanity will produce about 1 320 EB of digital data per year. This would amount to 10 % by limit $V_\infty$.

Behavior of the first and second derivatives of $V(T)$ values, shown in Figures 3 and 4, respectively. The maximum speed of growth of production volumes of digital data will be achieved by mid 2015 and will be 1 489.7 EB per year. In mid 2015, the second derivative changes sign. And will "inhibition" on growth of production volumes of digital data.

We calculate the value of the total amount a digital data ever produced by humanity. To do this we integrate expression (5):
$$I(T) = \int_{T_b - \Delta T_b}^{T} V(\tau) d\tau, \qquad (6)$$

where $\Delta T_b$ – fitting coefficient, $\Delta T_b \approx 1.33$.

The value of $V(T)$ is this also a lower estimate of the amount of data, which makes extensive use of humanity in their lives within one year. Then the expression:
$$K_I(T) = \frac{V(T)}{I(T)}, \qquad (7)$$

it makes sense to lower estimates of the activity coefficient of use digital data. Dependence of $K_I(T)$ is shown in Figure 5.

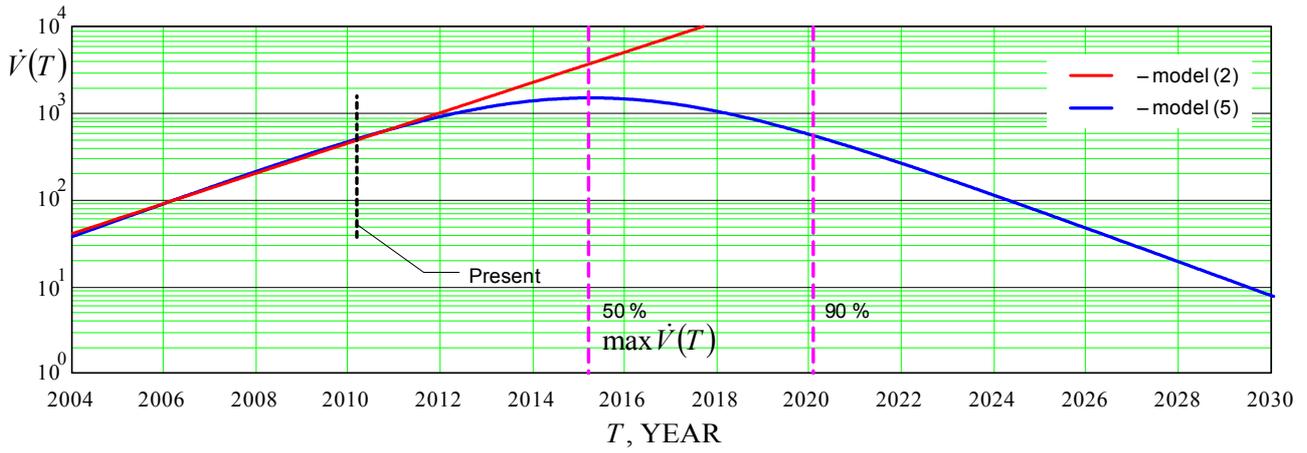

**Figure 3.** The first derivative values $\dot{V}(T)$, model (5)

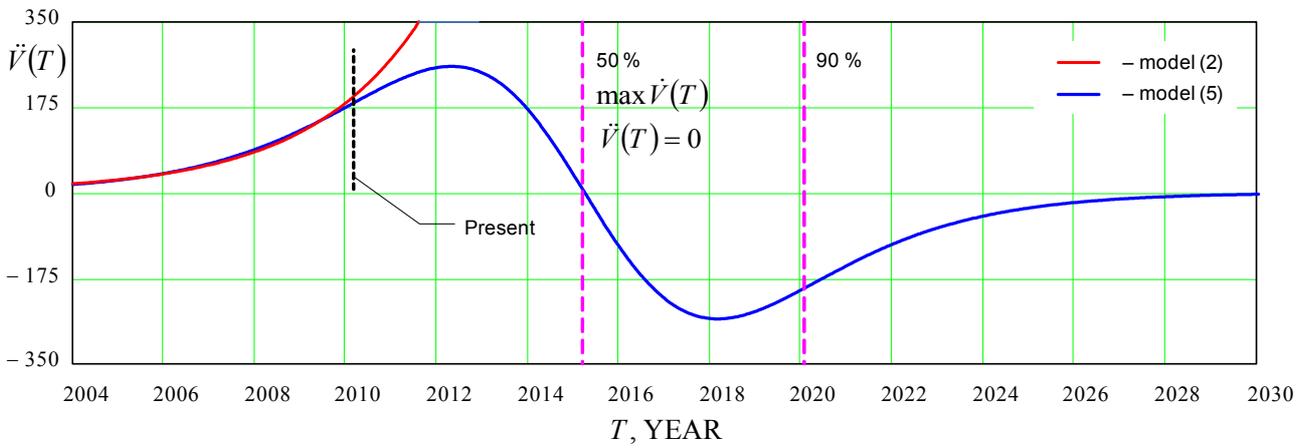

**Figure 4.** The second derivative values $\ddot{V}(T)$, model (5)

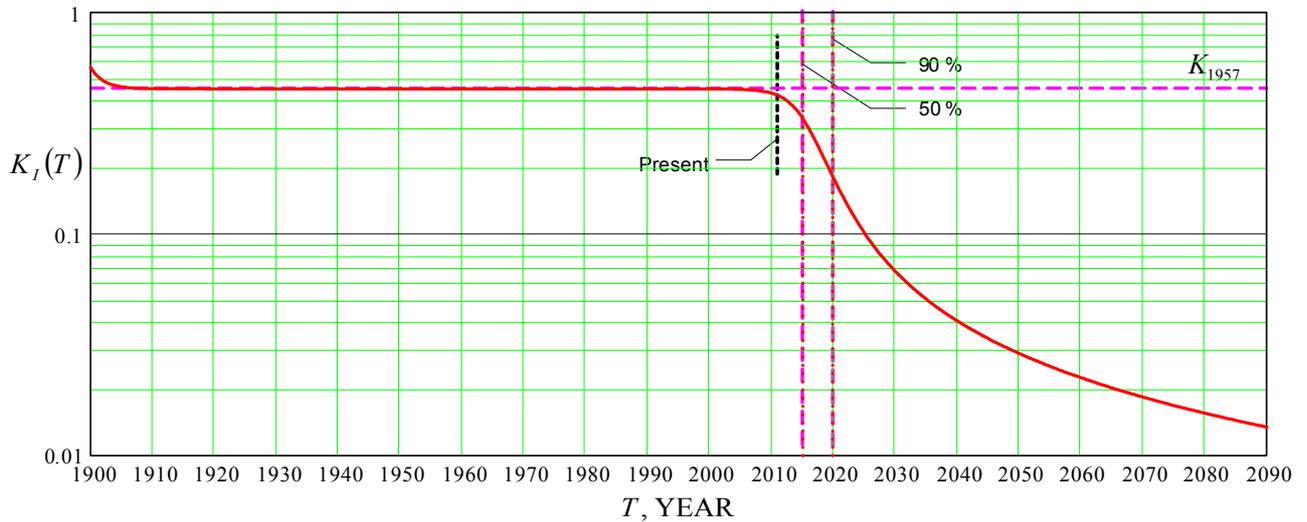

**Figure 5.** Activity coefficient of the use of digital data

From the properties of expressions (5) and (6) that the magnitude of the $K_I(T)$ monotonically depends on the $T$. And besides this:

$$\lim_{T \to T_b} K_I(T) = 1, \quad \lim_{T \to +\infty} K_I(T) = 0.$$

From the figure 5 that a long enough period of time (about 90 years), the activity coefficient was at a level close to $K_{1957} \approx 45.14$ %.

By the end of 2013, the total amount of digital data ever generated by humanity, will be $I \approx 9.137$ ZB. And coefficient of activity of its use will be at the level of $K_I \approx 38.77$ %.

Thus, a model (5) is adequately reflects the past and present. It's a generates in the principle, plausible estimates of $V(T)$. Nevertheless, the reasonable question arises: "Can we trust the number of 13.2 ZB?". To answer this question, we consider the dependence of the $V(T)$, in the aspect of the demographic situation on Earth.

## 4. Relationship of production a digital data with demographics of the Earth

Let us formulate a **postulate**: "*Population of planet Earth – is the main source and consumer of digital information produced on the planet Earth*". It's a quite confirmed by the results is studies [2, 12, 13]. Consequently, the behavior of should be considered in relation to the demographic situation on Earth. There is a theory which satisfactorily describes the population dynamics of the Earth [14]. The dependence of $N(T)$ – the number of people on Earth, on the time $T$ is written as:

$$N(T) = \frac{c_H}{\tau_H} \operatorname{arcctg} \frac{T_0^H - T}{\tau_H}, \qquad (8)$$

where the values of the constants defined with an error of a few percent, are:

$$c_H = (186 \pm 1) \cdot 10^9; \quad T_0^H = 2007 \pm 1; \quad \tau_H = 42 \pm 1.$$

Note that the values (8) with sufficient accuracy coincide with the data of UNO [15] and IIASA [16]. The dependence of the $N(T)$ for the period $T = 1900 \ldots 2090$ years, is shown in Figure 6. In the same time scale in Figure 7 shows the dependence of $V(T)$.

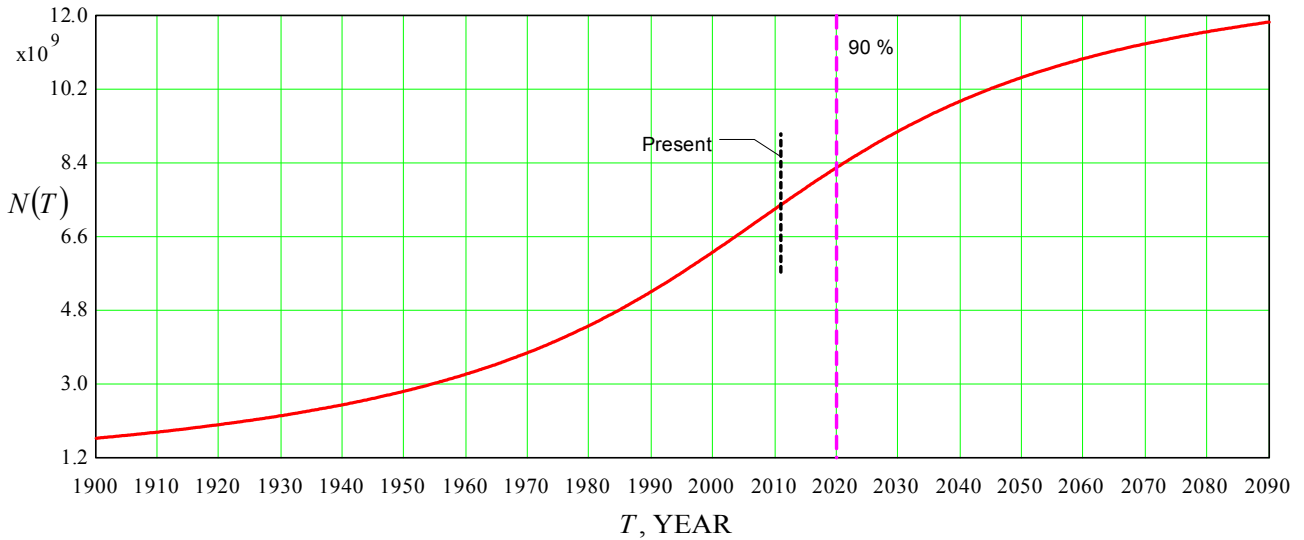

**Figure 6.** Population size of Earth, billions of people

Take the ratio of $V(T)$ and $N(T)$:

$$R(T) = 8 \cdot 10^{18} \frac{V(T)}{N(T)} \cdot \frac{10^{-3}}{365 \cdot 24 \cdot 3600}. \qquad (9)$$

This ratio characterizes the (average) effective amount of digital data in kbps, which produces one person per second. Dependence (9) is shown in Figure 8. In a logarithmic scale, the dependence of $R(T)$ is also shown in Figure 9.

Thus, the value of $R(T)$ has two characteristic features (see Figure 8): maximum $R_{max} \approx 379.72$ Kbit/s – it falls due in the middle of 2023, and asymptotic value $R_\infty \approx 241.17$ Kbit/s. Figure 9 shows that since the beginning of the digital era, and up to 2010, inclusive, the dependence of the $R(T)$ was very close to exponential.

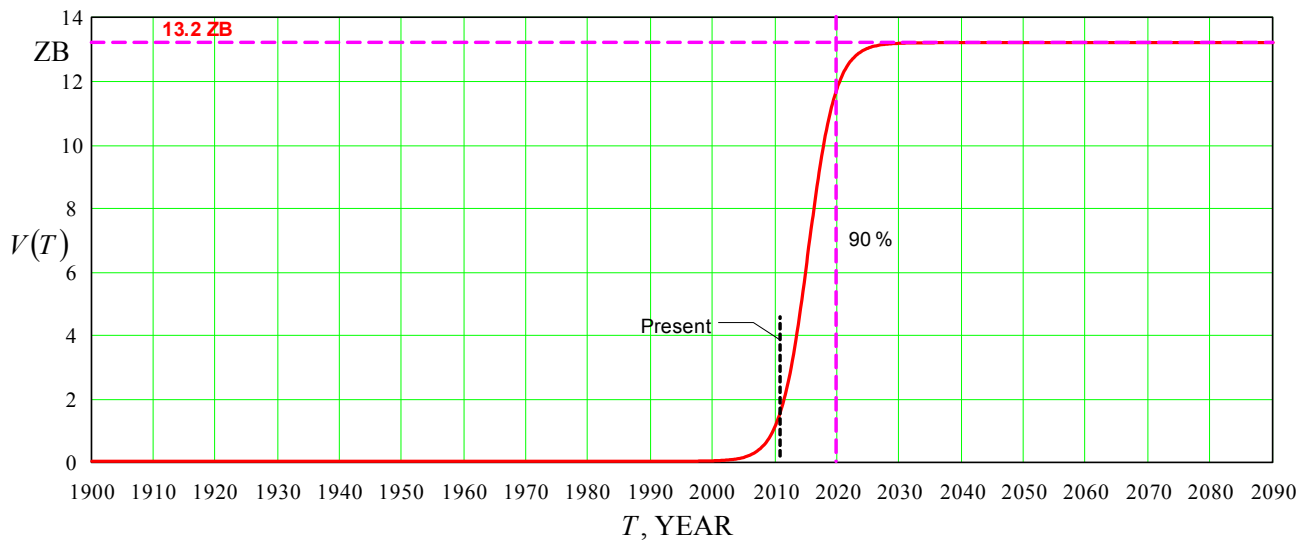

**Figure 7.** Production volumes of a digital data for per year, ZB

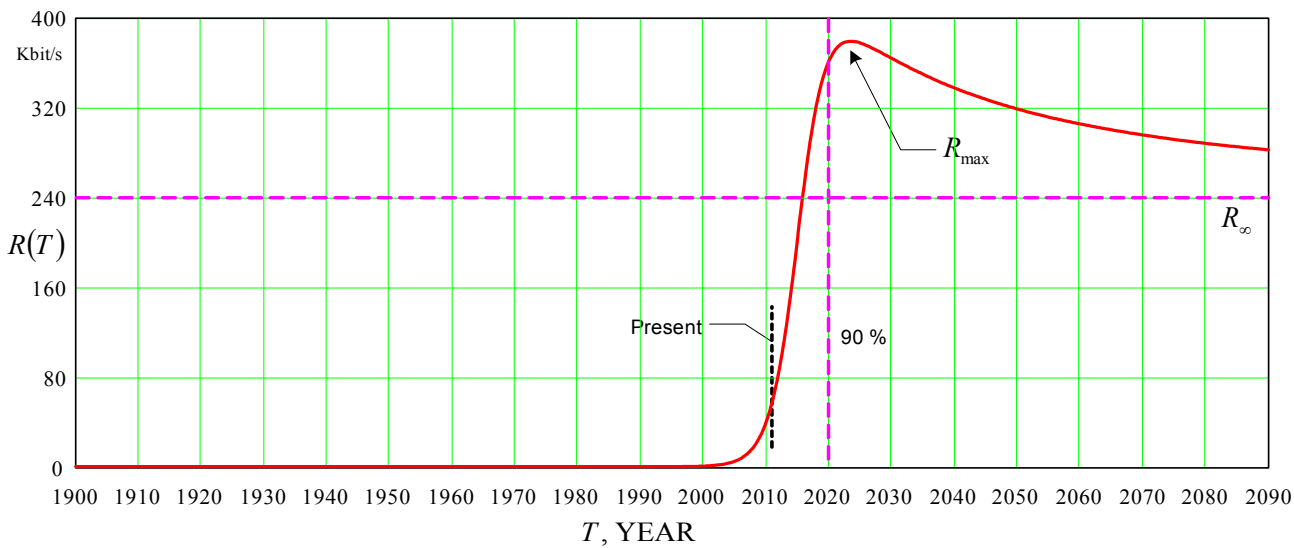

**Figure 8.** The effective rate of generation of a digital data by one man, Kbps

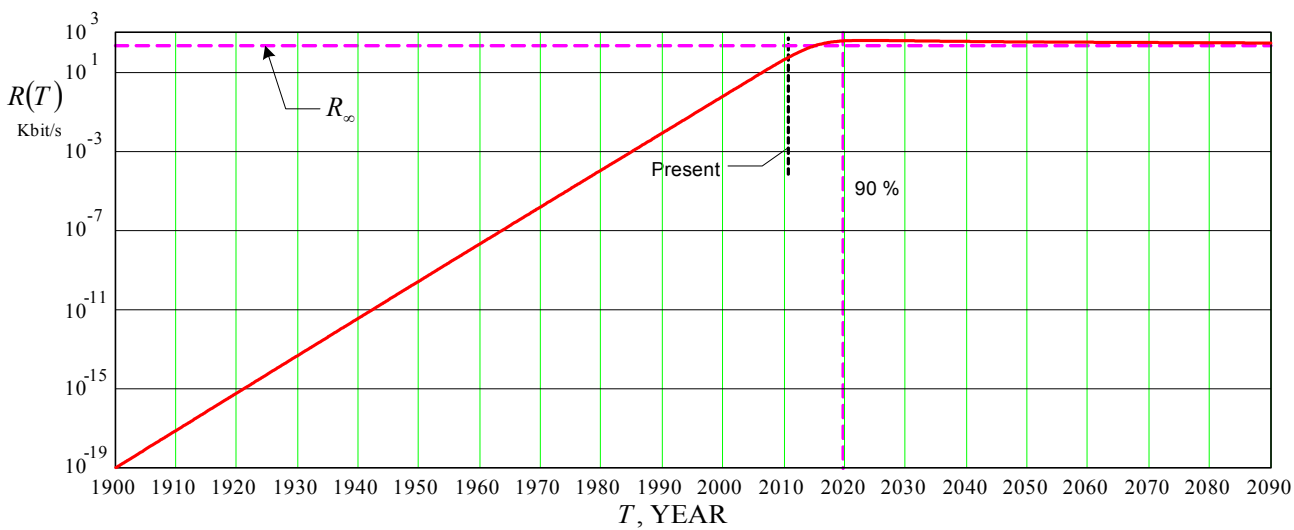

**Figure 9.** The effective rate of generation of a digital data by one man, Kbps

We calculate the total number of people who lived on the planet since the beginning of a digital era. To do this we integrate expression (8):

$$P(T) = \int_{T_b^H}^{T} N(\tau)\,d\tau, \quad T_b^H = T_b - \Delta T_b - \frac{\tau_H}{2}. \quad (10)$$

Construct the ratio:

$$Q_I(T) = \frac{I(T)}{P(T)} \cdot \frac{10^{18}}{10^9}, \quad (11)$$

that describes how many digital data (on average) accounts for one person, in GB. This dependency is shown in Figure 10. The value of $Q_I(T)$ in the future tends to the asymptotic value $Q_\infty \approx 948.88$ GB. By the end of 2011, for each inhabitant of the Earth will represent about 20.53 GB of digital data.

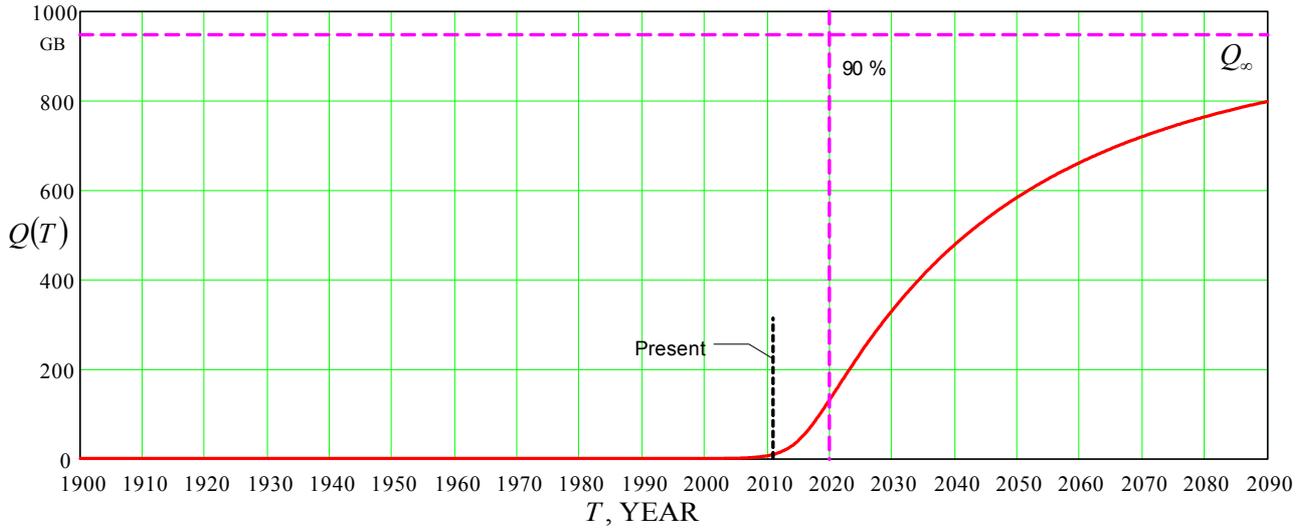

**Figure 10.** The volume of a digital data attributable to each inhabitant of the Earth, GB

Construct the ratio (volume of digital data fall on each inhabitant of the Earth divided by the effective rate of generation of digital data by one person):

$$\tau_I(T) = \frac{10^9 \cdot 8}{10^3} \cdot \frac{Q(T)}{R(T)} \cdot \frac{1}{3600 \cdot 24}, \quad (12)$$

has dimension of time and expressed in days. Dependence $\tau_I(T)$ is shown in Figure 11.

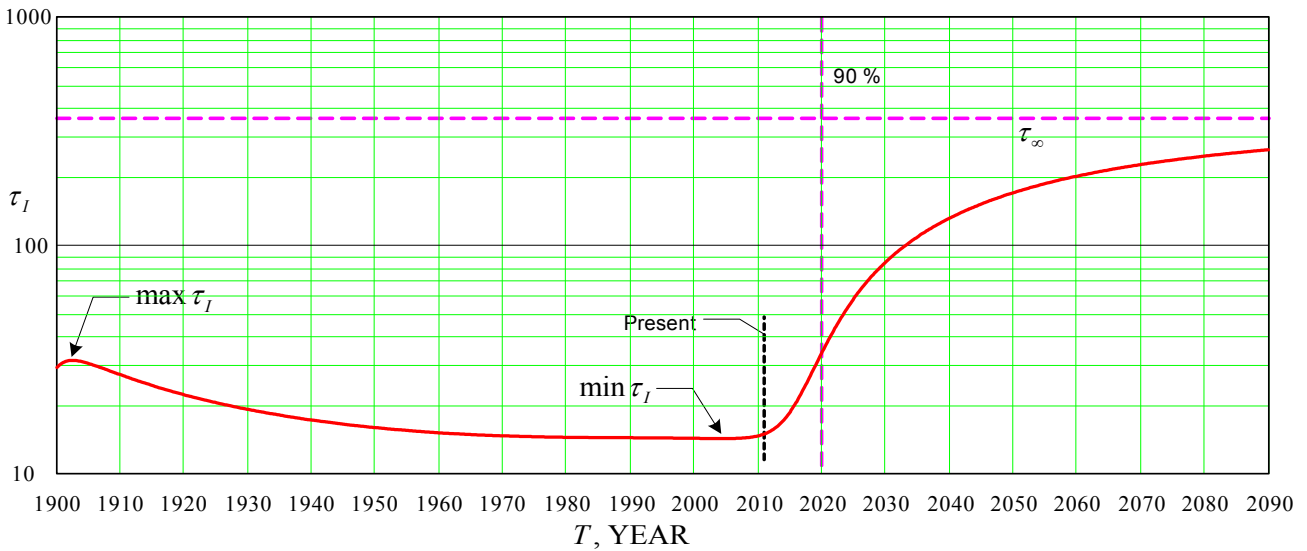

**Figure 11.** The effective time of generating a digital data inhabitant of Earth, Day

The value of $\tau_I(T)$ characterizes the effective duration of generating digital data from one inhabitant of Earth. At the beginning of the digital era $T_b$ (late 1897 года), the value of $\tau_I$ was about 2.5 weeks. Then its value rose to 4.5 weeks – in mid 1902 ($\max \tau_I$), and then started to decline. Minimum value $\tau_I$ – about 2 weeks ($\min \tau_I$), came in mid 2004.

Calculations show that in the future, the dependence of the $\tau_I(T)$ tends to an asymptotic value of $\tau_\infty \approx 1$ year. In late 2013 the effective time of generating digital data by one person will be about 16 days.

## 5. Conclusion

This paper presents a phenomenological model the growth production volumes of digital data. Model differs by scenarios of explosive and exponential growth. It's a show that in the future, the growth volume of digital data stabilize at 13.2 ZB per year. In this case, the level of 50 %, from the limiting value $V_\infty$, humanity will reach by mid 2015. Level 90 % – by 2020. By mid 2011, humanity will produce about 1 320 EB of digital data per year. This would amount to 10 % by limit $V_\infty$. By the end of 2013, the total amount of digital data ever generated by humanity, will be $I \approx 9.137$ ZB.

Thus, the passage by $0.1 V_\infty$ (2011) to $0.9 V_\infty$ (2020) happen in approximately for 9.5 years. The data of Figure 7 show that the period 2000-2030 is rightly describe as the digital revolution.

From the Figure 5 shows that during the whole XX century of the activity coefficient of use digital data was below 0.5. More than half of digital data created of humanity, later not been claimed. Hereinafter the situation will only worsen. In 2025, the world will actively use no more than 10% of the created digital data. Such low values of the $K_I$ will strengthen the role of systems of structure-forming information, and extracting specific knowledge, facts and meaning out of data.

Three points to which there corresponds the model, lead us to believe with some certainty about its adequacy and accuracy.

<u>First,</u> the model has a high value of Pearson's correlation coefficient with reference data, $\mathrm{Corr}(V(T), \hat{V}) = 0.9999361$. The data cover the period of 2006-2012 years and are taken from sources [1-3]

<u>Secondly,</u> the model passes test on the prediction of the "backward". It generates an estimate of $T_b$ (early the digital era) – roughly the second half of 1897. Naive averaging given dates gives the result – 1883 year. The averages weighted model errors (relative to "historical date") are: $\Delta_{Tb} \approx 7.39$ year (absolute) и $\Delta^*_{Tb} \approx 6.829$ % (relative).

<u>Third,</u> the simulation results correlate well with the demographic situation on Earth. Comparison of these two categories is quite correct (if we start from the postulate formulated in Section 4 of this article).

The effective rate generation of the digital data given on a single inhabitant of Earth, the end of 2013 amount to 118.47 Kbit/s (value calculated by the model). According to IDC [17] by the end of 2013, 28.7 % of the world's population will have access to the Internet, and only 4.6 % – broadband access. According to the company Cisco [12], in 2013, the annual volume of Internet traffic will be 388.332 EB. (consumer segment). This is 10.96 % of the total amount of digital data generated of humanity in 2013 (value calculated by the model). Based on these values may calculate the effective rate of generation of digital data on the Internet (normalized on a single Internet user) in 2013:

$$R^{2013}_{UserNet} = \frac{R(2013)}{0.046} \cdot \frac{388.332}{V(2013)} = 282.348 \text{ Kbit/s.} \qquad (13)$$

The normalization is performed on the number of people on Earth who have broadband Internet. Since this segment is considered in white papers Cisco [11, 12].

In 2013, the total time spent by users on the Internet amount to 6,776 trillion seconds per month [13]. Averaged on a single user traffic will be 4.86 MB per minute. Based on these values is possible calculate an estimate similar to (13):

$$\hat{R}_{UserNet}^{2013} = \frac{6.776 \cdot 10^{12}}{N(2013) \cdot 0.046} \cdot \frac{1}{30 \cdot 24 \cdot 60} \cdot \frac{4.86 \cdot 8 \cdot 10^3}{60} = 291.6 \text{ Kbit/s.} \quad (14)$$

Relative average-weighted deviation between the values of $R_{UserNet}^{2013}$ and $\hat{R}_{UserNet}^{2013}$ is 3.224 %. This indicates a very precise value of $R(T)$, which forms the a model (5). It is therefore possible to trust both the dependence of $V(T)$, so and limiting value $V_\infty$. Certainly within the error of initial the data $\hat{V}$ [1-3].

As shown in Figure 8, in the future are stabilized both quantities: $V(T)$ and $R(T)$, $V_\infty \approx 13.2$ ZB, $R_\infty \approx 241.17$ Kbit/s. The second value will pass through a maximum $R_{max} \approx 379.72$ Kbit/s – in mid 2023. This fact has the a logical explanation. In its basis again lies a postulate, formulated by in Section 4 of this article.

In the future world population will stabilize (estimation for the near future: 12.5 billions of people [14, 15]), literacy will become global (According to the UN, at the end of 2010, the literacy rate is has already 83 % [18]). Demographic growth potential of the data traffic would exhausted. On the other side: communication will become global; video telephony will widespread; will into use 3D-TV; and other broadband services [13]. Perhaps a significant proportion of traffic will be generated at information communications between computers. Hers appearance is predicted in the forecasts, including, and Good I. J. [4], and Ray Kurzweil [5]. But do not forget about other processes. Which are, in turn, reduces the global data traffic. They include:

- improving the efficiency of data compression;
- improving the depth and quality extraction of knowledge, facts and meaning out of data;
- structuring and normalization of data storage;
- increase of the share and depth of local intellectual data processing.

Worthy of attention also the data in Figure 11. The minimum value $\tau_I$ – about 2 weeks ($\min \tau_I$), came in mid 2004 year. Very close to this time lies the value of $T_0 = 2002$ year – when the volume of digital data has exceeded the volume of analog data [2]. Most likely, it is not an accidental coincidence. And if in 2013 each person will spend on generating digital data 16 effective days, then in the distant future for this he will have to expend one effective year.

Of course, numeric values that forms the model are not absolute. They will be adjusted in the future, along the measure updating the forecasts which given in [1-3, 12, 13]. But the basic trend is likely to continue. Unless of course on simulated system will not be exerted external and unaccounted influence.

An a small comment. Supporters of the technological singularity [5, 19], called the timing of its occurrence in the period 2020-2030. Let's look at the charts on the figures 7-9. Follows from them that, indeed, after 2030, mankind is likely to will acquire new quality of informatization society. But no collapse, explosion, they are do not portend. Because in the basis of civilization lies the logistic function.